\documentclass[fleqn,twoside]{article}
\usepackage{espcrc2}
\usepackage{graphicx}
\usepackage{epsfig}
\usepackage[figuresright]{rotating}

\newcommand{\AmS}{{\protect\the\textfont2
  A\kern-.1667em\lower.5ex\hbox{M}\kern-.125emS}}

\title{Final results on the neutrino magnetic moment from the MUNU experiment}

\author{The MUNU collaboration: 
Z.~Daraktchieva \address[NE] {Institut de physique, A.-L. Breguet 1, 
                   CH-2000 Neuch\^atel, Switzerland},
C.~Amsler \address[ZH]{Physik-Institut, Winterthurerstr. 190, 
CH-8057 Zurich, Switzerland},
M.~Avenier \address[GR]{Laboratoire de Physique Subatomique et de 
Cosmologie, IN2P3/CNRS-UJF, 53 Avenue des Martyrs, F-38026 Grenoble, France}, 
C.~Broggini \address[PD]{INFN, Via Marzolo 8, I-35131 Padova, Italy},
J.~Busto \addressmark[NE],
C.~Cerna \addressmark[PD],
F.~Juget \addressmark[NE],
D.H.~Koang \addressmark[GR],
J.~Lamblin \addressmark[GR],
D.~Lebrun \addressmark[GR],
O.~Link \addressmark[ZH],  
G.~Puglierin \addressmark[PD],
A.~Stutz \addressmark[GR],
A.~Tadsen \addressmark[PD],
J.-L.~Vuilleumier \addressmark[NE],
V.~Zacek \address[MO]{Universit\'e de Montr\'eal, C. P. 6128, Montr\'eal, P.Q.
Canada H3C 3J7}}
\begin{document}
\begin{abstract}
The MUNU detector was designed to study $\overline{\nu}_ee^{-}$
elastic scattering at low energy. The central component is a Time
Projection Chamber filled with CF$_4$ gas, surrounded by an
anti-Compton detector. The experiment was carried out at the Bugey
(France) nuclear reactor. In this paper we present the final analysis
of the data recorded at 3 bar and 1 bar pressure. Both the energy and
the scattering angle of the recoil electron are measured. From the
3 bar data a new upper limit on the neutrino magnetic moment
$\mu_{e}^{short}< 9\cdot 10^{-11}$ $\mu_B$ at 90 \% CL was derived. At 1 bar
electron tracks down to 150 keV were reconstructed, demonstrating 
the potentiality of the experimental technique for future applications 
in low energy neutrino physics.
\vspace{1pc}
\vspace{0.5cm}
\end{abstract}

\maketitle

\section{Introduction}

%\vspace*{-1.cm}
\begin{figure}[htb]
\begin{center}
\hspace*{-0.7cm}
\epsfig{file=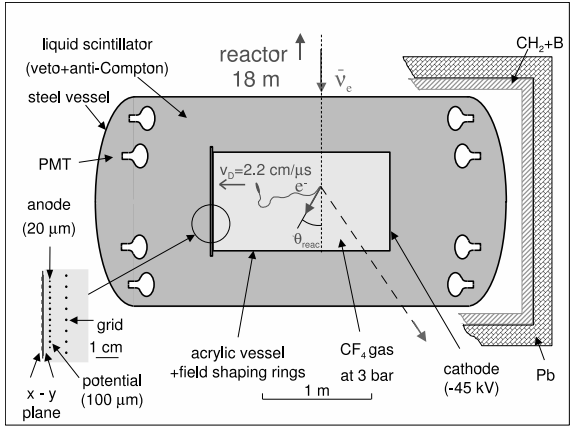,width=7.5cm}
\caption{The MUNU detector at the Bugey reactor.}  
\label{fi:MUNU_det}
\end{center}
\end{figure}
%\vspace*{-1.cm}
\begin{figure}[hbt]
\begin{center}
\vspace*{-1.cm}
\epsfig{file=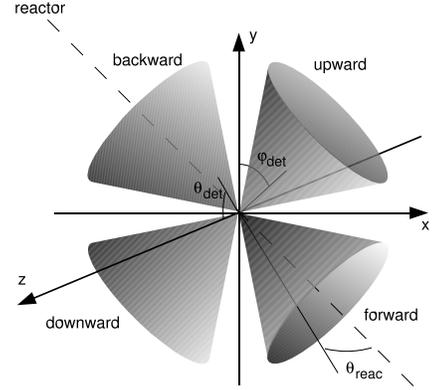, height=6.cm}
\caption{The four kinematical cones.}
\label{fi:cones}
%\vspace*{-1.cm}
\end{center}
\end{figure}

The MUNU experiment was designed to study $\overline{\nu}_ee^{-}$
elastic scattering at low energy and to probe the existence of a
magnetic moment of the electron antineutrino. The detector is located
at 18 m from the core of a 2.75 GWth commercial nuclear reactor in
Bugey (France). The central component is a CF$_4$ gas Time Projection
Chamber (TPC).  In \cite{MUNU3} we presented an analysis of 66.6 days
live time of reactor-on data, as well as 16.7 days of reactor-off
data, taken at a pressure of 3 bar. In this Letter we present the
final analysis, using the same data set, and the same event selection,
but taking better advantage of the electron kinematics, extending the
area in which the background is measured and achieving a more precise
determination. Moreover we present the analysis of 5.3 days live time
of reactor-on data taken at 1 bar pressure.

Technical details of the MUNU detector have already been presented in
ref.\cite{MUNU1,MUNU2}. Here we only describe the most essential 
features.

\section{The experiment}

The central part of the detector is a cylindrical TPC filled with 
CF$_4$ gas. Measurements were performed at a pressure of
3 bar (11.4 kg of CF$_4$) and 1 bar (3.8 kg). The gas serves as target and
detector medium for the recoil electrons. CF$_4$ was chosen because of
its high density, low atomic number, which reduces multiple
scattering, and its absence of free protons, which eliminates the
background from $\overline{\nu}_{e}p \rightarrow e^{+}n$. As shown in
fig. \ref{fi:MUNU_det} the gas is contained in a 1 m$^{3}$ acrylic
vessel 90 cm in diameter and 160 cm long. The drift volume is defined
on one end by a plain cathode and the other one by a grid made from
wires. An anode plane made of 20 $\mu$m wires with a pitch of 4.95 mm,
separated by 100 $\mu$m potential wires, is placed behind the grid to
collect and amplify the ionization charge. The integrated anode signal
gives the total deposited energy. An x - y read-out plane is located
behind the anode plane.  It contains x strips on one side, and
perpendicular y strips on the other one.  The pitch is 3.5 mm. The
signals induced in these strips provide the spatial information in the
x and y directions. The third projection z is obtained from the time
evolution of the signal. The drift field was selected such as to
achieve a drift velocity of 2.15 cm/$\mu$s.

It must be noted that the detector is installed under the reactor
core, at an angle of 45$^{\circ}$.  The detector axis is perpendicular
to the reactor core-detector axis.  The anode wires are parallel to
the reactor core-detector axis, and the grid wires perpendicular to
it. The x strips are vertical and the y strips horizontal. The
read-out plane is therefore by construction symmetric with respect to
four directions: reactor core to detector and opposite, as well as the
two orthogonal directions.

\begin{figure}[hbt]
\begin{center}
\vspace*{-1.cm}
\epsfig{file=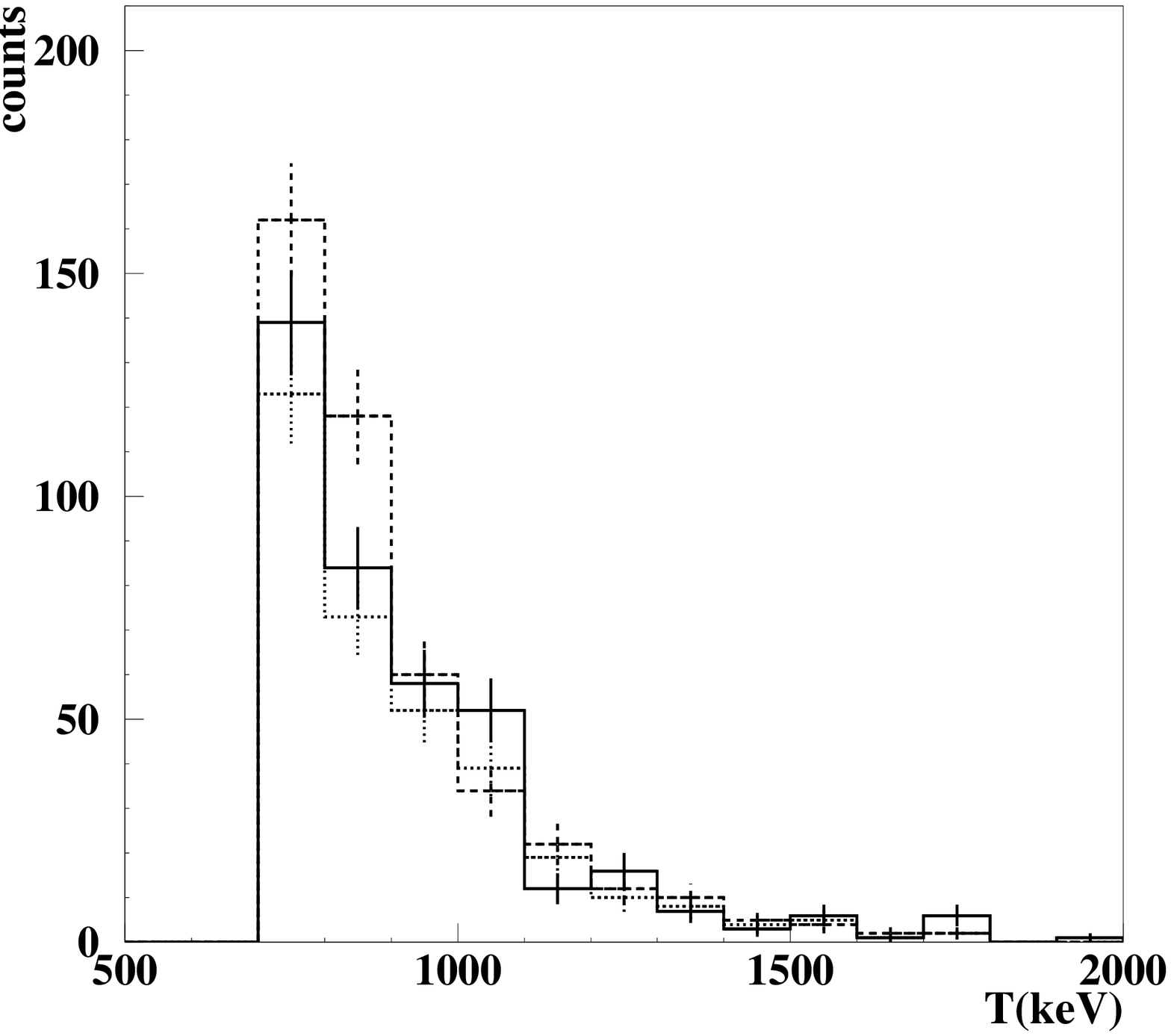, height=6.3cm}
\vspace*{-1.0cm}
\epsfig{file=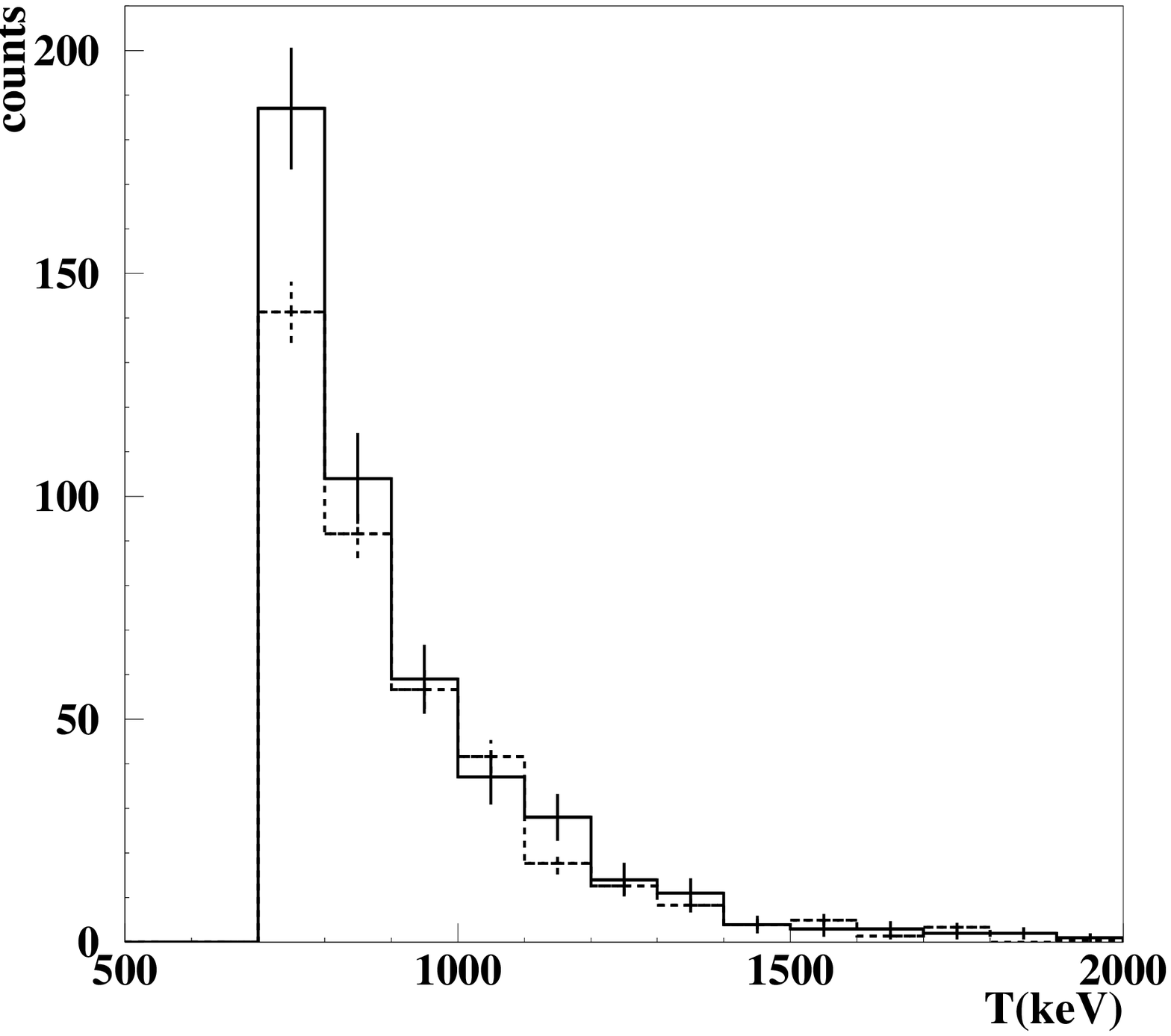, height=6.3cm}
\caption
{3 bar data, reactor-on, energy spectra; top: upward (dashed
line), downward (solid line) and backward (dotted line) electrons;
bottom: forward (solid line) and NB (dashed line) electrons}
\label{fi:udb}
%\vspace*{-1.cm}
\end{center}
\end{figure}
The acrylic vessel is immersed in a steel tank (2 m diameter and 3.8 m
long), filled with 10 m$^3$ of liquid scintillator (NE235) and viewed by
48 photomultipliers (PMT). The liquid scintillator acts as an
anti-Compton detector and as a veto against cosmic muons. 
The anti-Compton detector also sees the primary
scintillation light of heavily ionizing particles such as $\alpha$'s,
and the secondary light emitted during the amplification process
around the anode wires,
which provides a second measurement of the total deposited energy.  

In addition, the detector is shielded against local activities by 8 cm
of boron loaded polyethylene and 15 cm of Pb. The concrete and steel
overburden of the laboratory corresponds to the equivalent of 20 m of
water.

Events in the TPC not in coincidence with a signal above 22 MeV in the
scintillator within an 80 $\mu$s time window are recorded.  The
selection of good events off-line proceeds in two steps. First a software
filter rejects the muon related events, Compton electrons, discharges
and uncontained events. The selection is finalized in a visual scan.
A neutrino scattering candidate event is a
continuous single electron track fully contained in a 42 cm fiducial
radius, with no energy deposition above 90 keV in the anti-Compton in
the preceding 200 $\mu$s. The initial direction of the electron track
is obtained from a visual fit \cite{MUNU3}.  The scattering angles in
the x-z and y-z projections are determined first, and used to
calculate the scattering angle $\theta_{reac}$ with respect to the
reactor-detector axis, which coincides quite precisely with the
scattering angle, as well as the angle $\theta_{det}$ with respect to
the detector axis (see fig.\ref{fi:cones}). The angle $\varphi_{det}$
between the projection of the initial track direction on the
x-y plane and the vertical y axis is also determined.

As described in more details in \cite{MUNU3} we apply the
angular cut $\theta_{det} < 90^{\circ}$ to suppress the background
from activities on the read-out plane side of the TPC, which was found
to be noisier. This is presumably due to the greater complexity of the
anode side and the larger inactive volume in the scintillator. 

The reactor neutrino spectrum, necessary to interpret the data, was
calculated using the formalism described in \cite{MUNU3,Zac86}. Above
1.5 to 2 MeV neutrino spectra reconstructed from the measured $\beta$
spectra of the fission fragments were used. The uncertainty is of
order 5 \% or less in this energy range, in which the neutrino spectrum
was moreover thorougly probed in
measurements of $\overline{\nu}_{e}p\rightarrow e^{+}n$ scattering at
reactors. At lower energies calculations only are available.  We have
used the calculated neutrino spectra of the fission fragments from
\cite{BeaV99,VogE89}, and taken into account the neutron activation of
$^{238}$U, as discussed in \cite{Kop97}, which is significant below
1 MeV.  The uncertainty in the neutrino spectrum in this energy range 
was estimated to 20 \%.
\section{3-bar forward-normalized background analysis}
The 3 bar data were taken with a TPC trigger threshold of 300 keV.
Considering the entire event selection procedure, the live time,
limited primarily by the data transfer rate and the anti-Compton, was
65 \%.  At 3 bar the tracks are long enough to be scanned with
sufficient efficiency for electron kinetic energies
$T_{e}>$700 keV. As before we rely on
kinematics to select the good candidate events. For each electron
track the neutrino energy $E_{\nu}$ is reconstructed from the
scattering angle, taken as $\theta_{reac}$, and the measured electron
recoil energy $T_{e}$. Events with $E_{\nu}>0$ are declared forward
events since, effectively, this criteria selects electrons with an
initial track direction within a forward cone, the axis of which
coincides with the reactor core-detector axis. The opening angle
depends on the energy, and is larger than that of the kinematic cone
for recoil electrons ($E_{\nu}>T_e$). Simulations taking into account
the angular response of the detector show that nearly 100 \% of the
recoil electrons fall in the forward category, which however also
contains a contribution from the background, which is isotropic
around the detector axis.

To estimate the background the same procedure is applied in the
three directions equivalent, considering the read-out plane, to
the reactor core-detector direction, taken as reference for the
forward events. These directions are thus also equivalent from the
point of view of the angular response in $\varphi_{det}$, which is not
completely linear, and of the acceptance. As depicted in
fig.\ref{fi:cones} these directions define the backward cone, opposite
to the forward cone, and the two perpendicular upward and downward
cones \cite{MUNU4}. To avoid overlap of the cones, which can occur for
$T_{e} < 2m_{e}c^{2}$, we require in addition that the angle
$\varphi_{det}$ is within less than $45^{\circ}$ with respect to the
cone axis. This only reduces the acceptance of the forward cone for
recoil electrons in a negligible way.

While the forward electrons contain recoil plus background
events, the backward, upward and downward electrons are  background
events only. The energy distributions of the upward,downward and backward
electrons (1154$\pm$34 in total) are presented in fig. \ref{fi:udb}. The
distributions are compatible within the errors, which confirms the
isotropy of the background inside the TPC.  

\begin{figure}[htb]
\vspace*{-1.0cm}
\begin{center}
\epsfig{file=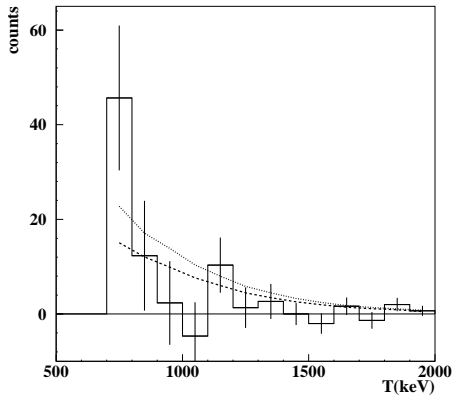, height=6.3cm}
\vspace*{-1.cm}
\caption{Energy distribution of the forward minus NB electrons above
700 keV, 3 bar, reactor-on; comparison with the expected spectrum for
weak interaction alone (dashed line) and for a magnetic moment of
$9\cdot 10^{-11}$~$\mu_{B}$ (dotted line).}
\label{fi:3bar_f-be}
\end{center}
\end{figure}

We normalize the background to the forward cone by dividing by 3 the
rates in the backward, upward and downward cones. This normalized
background (NB) is then directly compared with the event rate in the
forward cone.  The energy distributions of both forward (455$\pm$21)
and NB (384$\pm$11) electrons are shown in fig.\ref{fi:udb}. There is a clear
excess of forward events, from the reactor direction. The total number
of events forward minus NB above 700 keV is 71$\pm$23 counts for 66.6
days live time reactor-on, corresponding to 1.07$\pm$0.34 counts per
day (cpd). The forward minus NB spectrum representing the measured
electron recoil spectrum is displayed in fig. \ref{fi:3bar_f-be}.

\begin{figure}[hbt]
\begin{center}
\vspace*{-1.0cm}
\epsfig{file=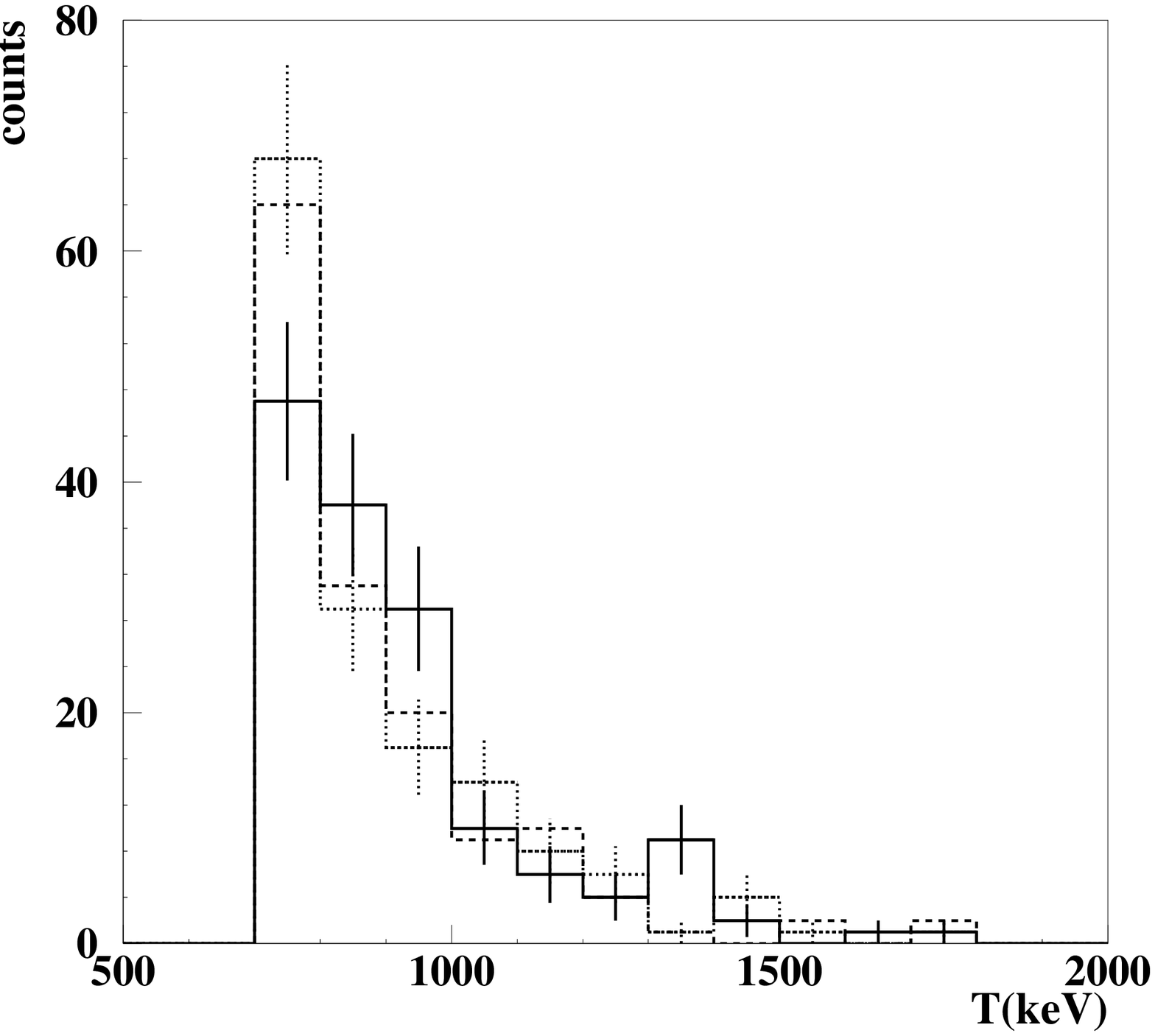, height=6.3cm}
\vspace*{-1.0cm}
\epsfig{file=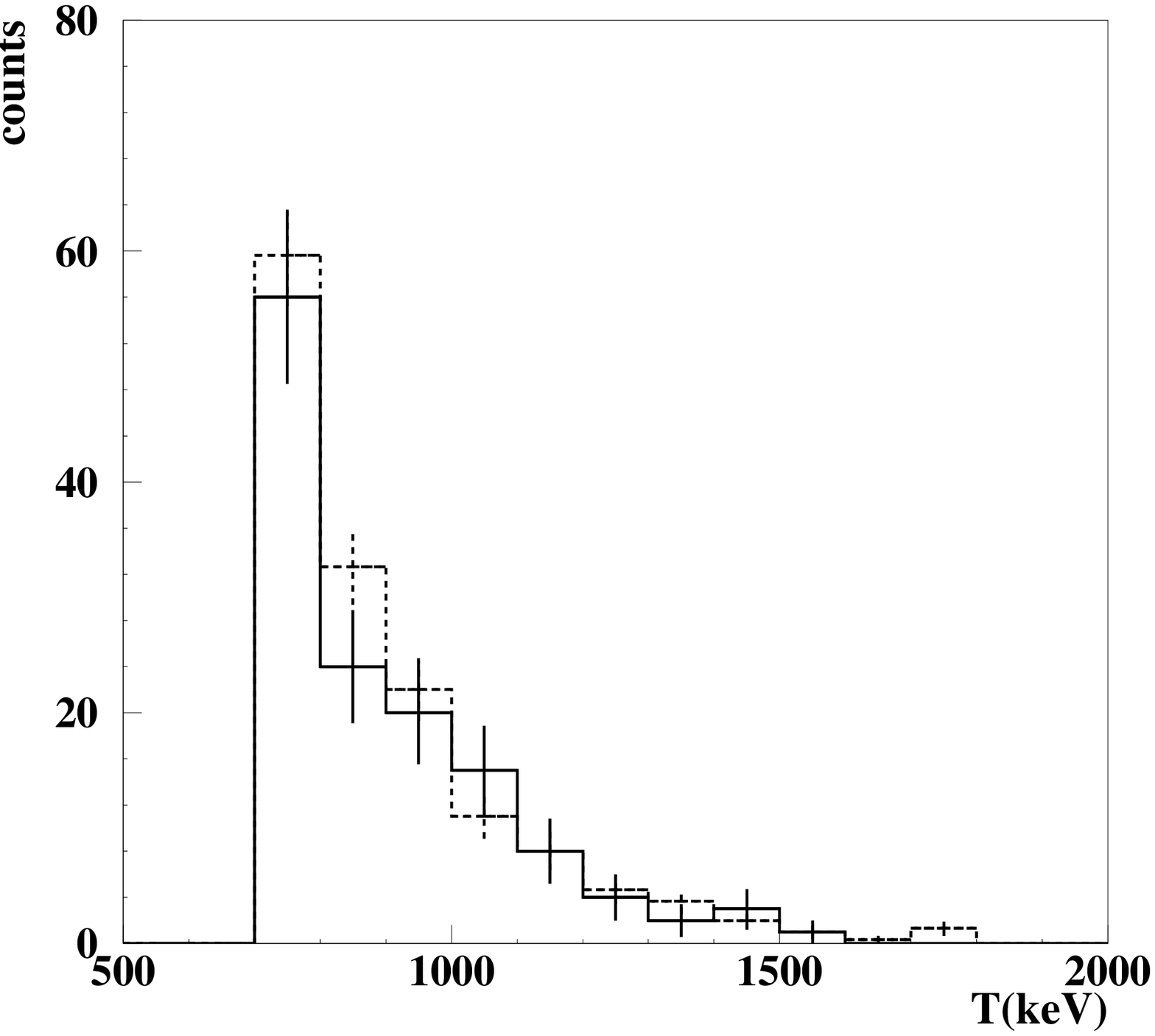, height=6.3cm}
\caption
{3 bar data, reactor-off, energy spectra; top: upward (dotted
line), downward (solid line) and backward (dashed line) electrons;
bottom: forward (solid line) and NB (dashed line) electrons}
\label{fi:udb_off}
%\vspace*{-1.cm}
\end{center}
\end{figure}

We make the same analysis with the data taken during the reactor-off
period as a cross check (16.7 days live time). The energy
distributions of both forward (133$\pm$11) and NB electrons
(147$\pm$7) are given in fig.\ref{fi:udb_off}.  The integrated
forward minus NB rate above 700 keV is -0.8$\pm$0.8 cpd, consistent
with zero.

%\vspace*{-1.0cm}
We now turn to the comparison with expectations. Monte Carlo
simulations (GEANT 3) were used to calculate the various acceptances
of the event selection procedure. The detector containment efficiency
in the 42 cm fiducial radius was found to vary from 63 \% at 700 keV,
50 \% at 1 MeV to 12 \% at 2 MeV, with a typical error of 2 \%. 
The errors on the global acceptance, including track
reconstruction efficiency (4 \%), are of order of 7 \%.  This leads
to an expected event rate above 700 keV of 1.02$\pm$0.10 cpd
assuming a vanishing magnetic moment. This is in good agreement with
the observed 1.07$\pm$0.34 cpd.

The measured and expected recoil spectra for no magnetic moment are
compared in fig.\ref{fi:3bar_f-be}. First we note that the large
excess of events in the first two bins (700- 900 keV) observed in our
previous analysis \cite{MUNU3} has to a large extend disappeared. It
is thus most likely due to a statistical fluctuation in the
background, more precisely determined in the present analysis. And
second the measured and expected spectra are seen to be in
agreement. 

For a more quantitative analysis the $\chi^{2}$ method was used as in
\cite{MUNU3}, with the same binning for Gaussian statistics to apply
(100 keV bins from 700 keV to 1400 keV, and then a bin from 1400 to
2000 keV). The magnetic moment is varied to find the best fit. We
remind that the neutrinos travel over a short distance only so that
the experiment probes the magnetic moment $\mu_{e}^{short}$, as
described in ref. \cite{MUNU3}, to be precise its square. The allowed
range at the 90~\% CL is $(\mu_{e}^{short})^{2}= (- 0.72 \pm
1.25)\cdot 10^{-20}$ $\mu_{B}^{2}$, with $\chi^{2}$=11.5 for 7 degrees
of freedom at the central value.  This result is compatible with a
vanishing
magnetic moment. To obtain a limit on $\mu_{e}^{short}$ we
renormalize to the physical region ($(\mu_{e}^{short})^{2}> 0$) and
obtain $\mu_{e}^{short} < 9(7) \cdot 10^{-11}$~$\mu_{B}$ at
90(68)~\% C.L.  This constitutes a small improvement over our previous
analysis \cite{MUNU3}, restricted to recoil energies above 900
keV. The improvement results from the lower threshold and the better
background estimation.

We can compare this with results from other experiments.  The TEXONO
collaboration is performing an experiment near the Kuo-Sheng reactor
in Taiwan \cite{TEXONO}. Thanks to the use of an Ultra Low Background
High Purity Germanium detector they achieve a very low threshold of
12 keV. The reactor-on minus reactor-off rates were found to be identical
and from that the limit $\mu_{e}^{short} < 1.3 \cdot 10^{-10}$ $\mu_{B}$ at
90 \% C.L was deduced.  Superkamiokande \cite{Kam99,{BeaV99}} reported the 
limit $\mu_{e}^{sol} < 1.5 \cdot
10^{-10}$ $\mu_{B}$ at 90 \% CL from the study of the shape of the 
recoil spectrum of electrons in the scattering of solar neutrinos.
Depending on oscillation
parameters, this quantity may however differ from $\mu_{e}^{short}$.

\section{1-bar forward-normalized background analysis}

The technical details of the 1 bar measurements will be presented
elsewhere.  Here we restrict ourselves to the most relevant
parameters. The trigger threshold on the electron recoil was lowered
to 100 keV. The trigger rate is 0.4 s$^{-1}$. The deadtime, here also
mostly due to the data read-out and transfer time, and to the
anti-Compton, is around 50 \%, somewhat less than for the measurements
at 3 bar. The energy calibration of the TPC is again determined with
$^{137}$Cs and $^{54}$Mn radioactive sources. The energy resolution is
found to vary from 10 \% (1 $\sigma$) at 200 keV to 6 \% at 600 keV
following an empirical $E^{0.57}$ law.  Data were accumulated during
5.3 days live time reactor-on.

\begin{figure}[htb]
%\vspace*{-1.0cm}
\begin{center}
\epsfig{file=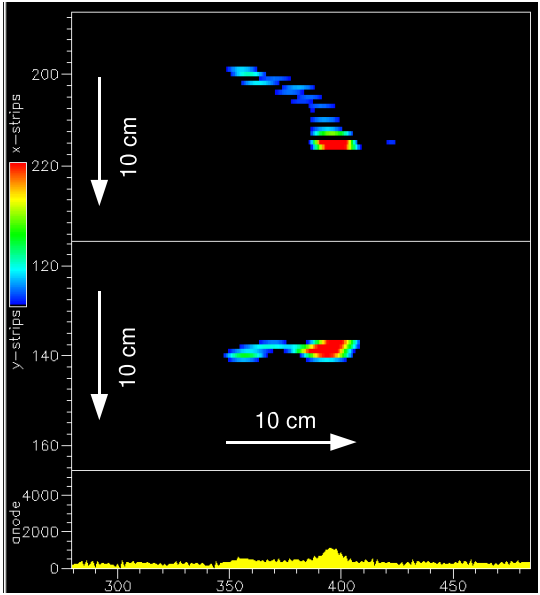, height=8.0cm}
\end{center}
\vspace*{-1.cm}
\caption{A 190 keV electron at 1 bar in the TPC; the xz, yz  projections are 
displayed as well as the anode signal.}  
\label{fi:1bar_el}
\end{figure}

The selection of good events proceeds in two steps, as with
the 3 bar data, ending with a visual scan. Tracks 
can be reconstructed with reasonable efficiency for energies above 150 keV.
Events above that energy are retained. 
As an example a 190 keV electron track in
1 bar of CF4 is shown in fig.\ref{fi:1bar_el}. The end of the track is easily
identified from the increased energy deposition (blob).  The
containment efficiency of the TPC in the 42 cm fiducial radius for
recoil electrons was calculated with the GEANT3 simulation code, and
found to vary from 85 \% at 200 keV, 50 \% at 400 keV to 5 \% at 1
MeV. The angular resolution was determined by scanning visually
simulated electron tracks. It varies from 15$^{\circ}$ (1 $\sigma$) at 200 keV,
12$^{\circ}$ at 400 keV, to 10$^{\circ}$ above 500 keV.

The
angular cut $\theta_{det}<$90$^{\circ}$ is also applied in the 1 bar
analysis.  The electrons are selected in
the same four kinematical cones. Here however the overlap of the cones is more
critical. The constraint in $\varphi_{det}$ is done somewhat
differently: downward electrons 190$^{\circ}$ to 270$^{\circ}$,
forward electrons 80$^{\circ}$ to 190$^{\circ}$, upward electrons
0$^{\circ}$ to 80$^{\circ}$ and backward electrons 270$^{\circ}$to
0$^{\circ}$.

\begin{figure}[hbt]
\begin{center}
\vspace*{-1.0cm}
\epsfig{file=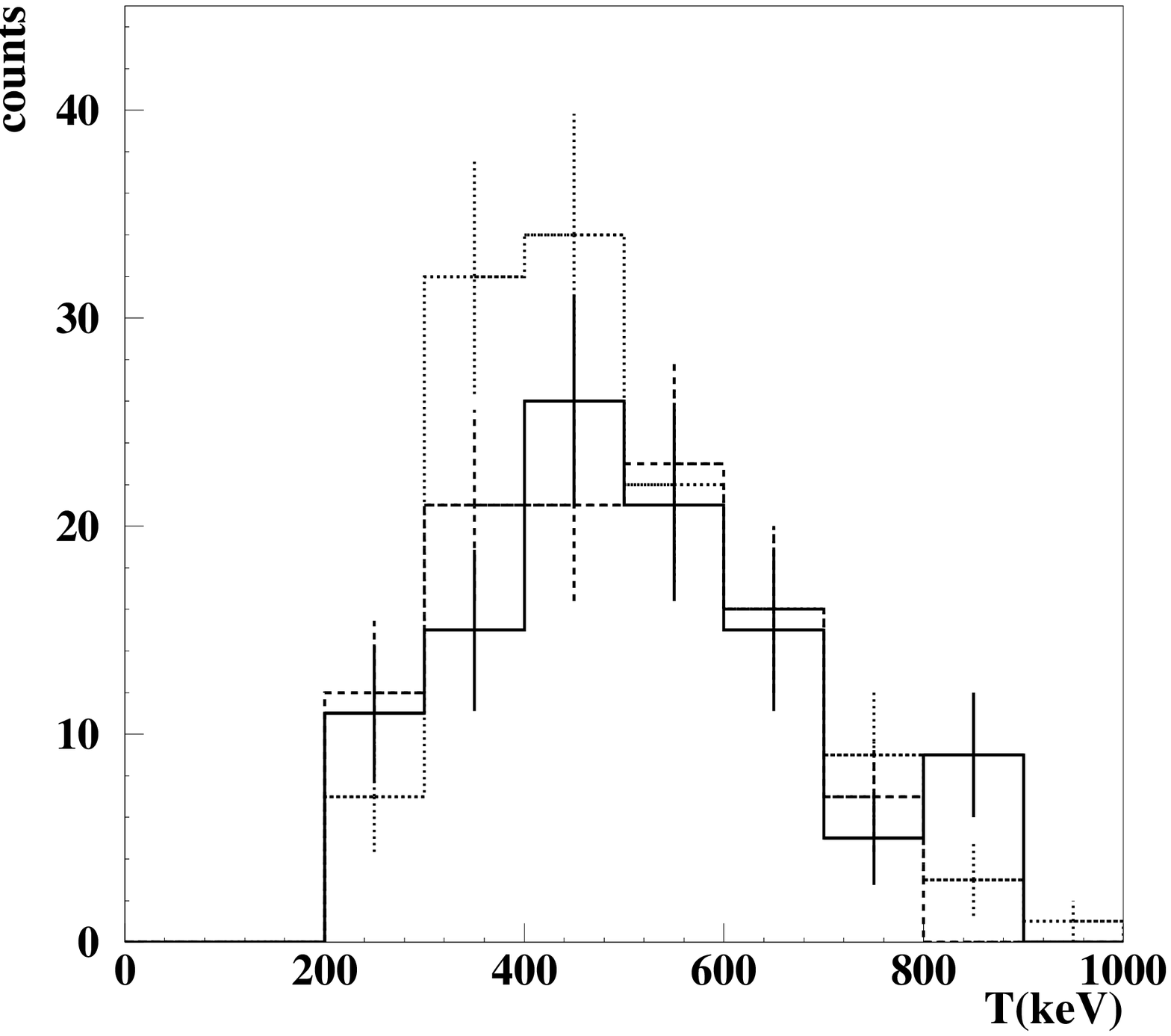, height=6.3cm}
\vspace*{-1.0cm}
\epsfig{file=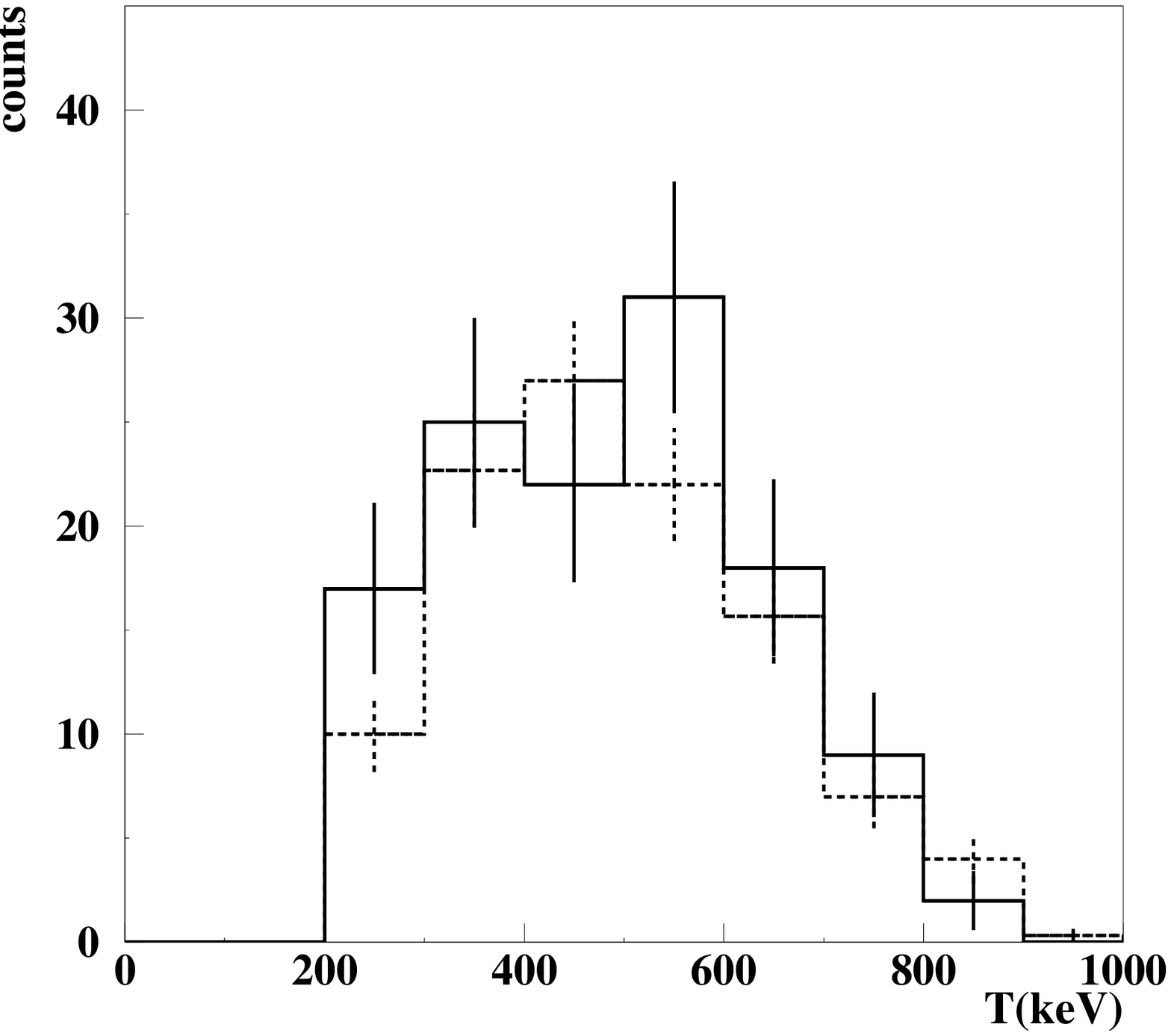, height=6.3cm}
\end{center}
\caption
{1 bar data, reactor-on, energy spectra; top: upward (dotted
line), downward (solid line) and backward (dashed line) electrons;
bottom: forward (solid line) and NB (dashed line) electrons}
\label{fi:1b_udb}
%\vspace*{-1.cm}
\end{figure}
This makes the solid angle for the forward electrons somewhat larger.
We expect from the Monte Carlo simulations that nearly 100 \% of the
recoil events above 200 keV will fall in it. The normalized background
NB in the three other cones has to be scaled accordingly, with a
factor which remains close to 3 however.  The energy distributions of
the upward, downward and backward electrons (326$\pm$18 in total) are
presented in fig.\ref{fi:1b_udb}.  They are seen to be compatible
within the errors.

\begin{figure}[htb]
\vspace*{-1.0cm}
\begin{center}
\epsfig{file=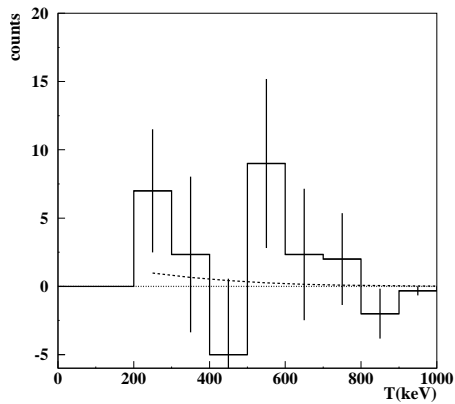, 
height=6.3cm}
\vspace*{-1.cm}
\caption{Energy distribution of the forward minus NB electrons above
200 keV, 1 bar, reactor-on; comparison with the expected spectrum for
weak interaction alone (dashed line).}
\label{fi:1b-f-nb}
\end{center}
\end{figure}

The energy distributions of the forward (124$\pm$11) and
NB (109$\pm$6) electrons are displayed in fig, \ref{fi:1b_udb}. The
difference is 15$\pm$12 events, giving an indication of a signal, 
corresponding to 2.89$\pm$2.39 counts per day. The energy distribution
is shown in fig.\ref{fi:1b-f-nb}. 
This measured total rate above 200 keV agrees with the expected
value 0.49$\pm$0.12 counts per day, assuming no magnetic moment.  The
error comes mainly from the uncertainties in the low energy part of
the neutrino spectrum, as described above. Due to the limited statistics
the data do not have a competitive
sensitivity to the neutrino magnetic moment. Nevertheless our
experiment shows that a gas TPC can be used to measure the energy and
direction of electrons with energies as low as 150 keV. Provided
background problems can be solved, this opens the way, for instance,
to on-line measurements of low energy solar neutrinos from the pp
reaction.
\section{Conclusions}
The MUNU experiment studied $\overline{\nu}_ee^-$ scattering at low
energy near a nuclear reactor, measuring both the energy and
scattering angle of the recoil electron. Thanks to a better estimation
of the background in a larger kinematical domain it was possible to
reduce the statistical uncertainties, analyzing data taken at 3 bar of
CF$_4$ during 66.6 days live time reactor-on with an energy
threshold of 700 keV.  Good agreement is observed between the measured
and expected recoil spectra assuming weak interaction alone. From this
the limit on the neutrino magnetic moment  $\mu_{e}^{short}<9 \cdot 
10^{-11}$ $\mu_{B}$ at 90 \% CL was derived.
Moreover reactor-on data were taken at 1 bar of CF$_4$ during 5.3 days 
live  time. Electron tracks were reconstructed efficiently down to 
150 keV recoil energy. This demonstrates the potentiality of gas TPC's for
possible future applications in low energy neutrino physics.

The authors want to thank the directors and the staff of EDF-CNPE
Bugey for the hospitality. This work was supported by IN2P3, INFN, and
SNF.

\end{document}